\documentclass[12pt]{article}

\usepackage{xcolor}
\usepackage{epsfig,latexsym,amsfonts,amsmath,amsthm,amssymb,amsbsy,multirow,slashed,wasysym,textcomp,subfigure,hyperref,wrapfig,datetime,comment,mathtools,cancel,cite}
\usepackage[font={footnotesize,sl},bf]{caption}

\def\ee{\end{equation}}
\def\be{\begin{equation}}
\def\bea{\begin{eqnarray}}
\def\eea{\end{eqnarray}}
\newcommand{\beq}{\begin{eqnarray}}
\newcommand{\eqq}{\end{eqnarray}}
 \newcommand{\badat}{\begin{alignedat}}
 \newcommand{\eadat}{\end{alignedat}}

\newcommand{\eal}[1]{\be \begin{aligned} #1 \end{aligned}\end{equation}} 
\newcommand{\eqn}[1]{\be #1 \end{equation}} 
\newcommand{\eqa}[1]{\bea  #1\end{eqnarray}}

\long\def\new#1\endnew{{\bf #1}}		
\long\def\del#1\enddel{}

\def\del{\partial}

\usepackage{color}

\newcommand{\pink}[1]{\textcolor{\pink}{#1}}

\definecolor{dblue}{rgb}{0.2,0.50,0.80}

\newcommand{\p}{\partial}

\def\ta{{\tilde \a}}
\def\E{{Q_4}}

\def\l{\lambda }

\def\a{\alpha }

\numberwithin{equation}{section} 

\begin{document}
\begin{titlepage}
  \thispagestyle{empty}
  \begin{flushright}
  \today
    \end{flushright}
  \bigskip
  \begin{center}
	 \vskip2cm
  \baselineskip=13pt {\LARGE \scshape{Hyperbolic  Vacua \\\vspace{0.5em} in Minkowski Space}}

	 \vskip2cm
   \centerline{Walker Melton, Filip Niewinski, Andrew Strominger, and Tianli Wang}
 \vskip.5cm
\noindent{\em Jefferson Laboratory, Harvard University,}
{\em Cambridge, MA, USA}
\bigskip
  \vskip1cm
  \end{center}
  \begin{abstract}
  Families of Lorentz, but not Poincare, invariant vacua are constructed for a massless scalar field in 4D Minkowski space. These are generalizations of the Rindler vacuum with a larger symmetry group. Explicit expressions are given as squeezed excitations of the Poincare vacuum. The effective reduced vacua  on the 3D hyperbolic de Sitter slices are the well-known de Sitter $\a$-vacua with antipodal singularities in the Wightman function. Several special interesting cases are discussed.  
  \end{abstract}
%

%
\end{titlepage}
\tableofcontents
\section{Introduction}
The $SO(3,1)\rtimes{\mathbb R}^{3,1}$ invariant Poincare vacuum for quantum field theory in Minkowski space ($M_4$) is characterized by the fact that inertial observers detect no particles. 
The Rindler vacuum on the other hand is characterized by the fact that observers uniformly accelerated away from a chosen plane detect no particles. This has a smaller $(SO(1,1)\times U(1))\rtimes{\mathbb R}^{2}$ symmetry. The relation between these two vacua is very rich and much  has been learned about quantum field theory (QFT), quantum entanglement, and black holes from studies of Rindler space. 

In this paper we construct  a new class of `hyperbolic' $M_4$ vacua for massless scalars with the full $SO(3,1)$ Lorentz symmetry but no translation symmetry. These vacua can be written as squeezed excitations of the Poincare vacuum (and vice versa). $M_4$ has a natural Lorentz-invariant foliation by hyperbolic slices. The region $S_4$ spacelike from the origin is foliated by $dS_3$ slices \cite{deBoer:2003vf}.  QFT on $dS_3$  has a well known  one-parameter family of $SO(3,1)$ invariant  `$\a$-vacua', of which the usual Euclidean vacuum is a special case \cite{Mottola:1984,Allen:1985,deBoer:2004nd}. The hyperbolic $M_4$ vacua can be understood as various uplifts of the $dS_3$ $\a$-vacua to $M_4$. The Euclidean vacuum on $dS_3$ uplifts to the Poincare vacuum on $M_4$. 

The construction of the hyperbolic vacua involves a Lorentz-invariant $Z_2$ action which mixes creation and annihilation operators. Important examples of these $Z_2$ are $PT$, conformal inversions (or shadows \cite{Chen:2023tvj,Jorstad:2023ajr}) combined with $PT$, and dilations by $i\pi$. 

Among the $dS_3$ $\a$-vacua is one with a special ($\mathrm{dS}_3$-mass-dependent) value of $\a$ for which the in/out Bogolyubov transformation is trivial and particle production is absent\cite{Bousso:2001mw}. In a conformal basis, annihilation or creation operator wave functions defining the uplift of this vacuum to 
$M_4$ vanish either above or below a null hypersurface passing through the origin.  We refer to this state  as the hyperbolic Rindler vacuum.  

Scattering amplitudes in QFT are traditionally studied in a basis which diagonalizes translations, so an expansion around hyperbolic vacua in such a basis  would not be useful. On the other hand, celestial holography typically employs a conformal basis \cite{Pasterski:2016qvg,Pasterski:2017kqt}, whose realized symmetries are not broken by hyperbolic vacua. Breaking four-dimensional translation invariance typically  results in  non-distributional celestial amplitudes that more closely resemble familiar two-dimensional conformal field theories \cite{Casali:2022fro,Ball:2023ukj,Stieberger:2022zyk,Fan:2022elem,Costello:2022jpg,Costello:2023hmi,Adamo:2023zeh,Melton:2022fsf,Gonzo:2022tjm,Stieberger:2023fju,Bittleston:2023bzp}. Hence celestial amplitudes in a hyperbolic vacuum  may provide a useful starting point for studying celestial conformal field theory. However in this paper we largely restrict to a bulk QFT analysis and do not elaborate on this application. 

We do not address herein stability/backreaction issues which arise beyond free field from interactions. We consider only massless scalars but expect our results to apply to any free massless field. 

This paper is organized as follows. After reviewing notations and preliminaries in Section 2, we define one class of hyperbolic vacua in Section 3 as  states annihilated by a linear combination of a negative frequency annihilation operator and its positive frequency $PT$ transform. Section 4 contains a similar construction combining the shadow transform with $PT$. We employ the recent realization \cite{Chen:2023tvj,Jorstad:2023ajr} of the shadow as a conformal inversion or reflection of the Einstein cylinder appearing in the conformal completion of Minkowski space to get a tractable expression for the Wightman function.    Section 5 generalizes this to construction to vacua where the positive-frequency solutions transform depending on the dilation weight of the free field. Section 6 identifies the hyperbolic Rindler vacua as a special one of these.

\section{Preliminaries}

This section gives our conventions and collects a few relevant formulae.  

Conformal primary wave functions of a massless field are given by 
\bea\label{cps}
 \phi^{\lambda,\pm}_q(X) &=& \frac{(\mp i)^{1+i\lambda}~\Gamma(1+i\lambda)}{(-q(z)\cdot X_\pm)^{1+i\lambda}} ,\cr  \text{with} \quad q^\mu(z)&=&(1+z\bar{z},z+\bar{z},-i(z-\bar{z}),1-z\bar{z}),
\eea
where $q^\mu$ is a future directed null vector, and $X^\mu_{\pm} = X^\mu \pm i \epsilon V^\mu$ with $V^\mu = (-1,0,0,0)$. These provide a complete basis of normalizable wave packets for $\lambda \in \mathbb{R}$. The overall normalization here is fixed so that they are related by simple Mellin transforms to plane waves. We specify the branch cut convention by
\be (x \pm i\epsilon)^{-1 - i\lambda}= \Theta(x)x^{-1-i\lambda}-\Theta(-x)e^{\pm \pi \lambda}(-x)^{-1-i\lambda}. \ee
An alternate basis is given  the shadow modes \cite{Pasterski:2017kqt}
\bea
 \widetilde{\phi}^{\lambda,\pm}_q(X)&=&\frac{(\mp i)^{1+i\lambda}~\Gamma(1-i\lambda)(X^2_\pm)^{i\lambda}}{(-q\cdot X_\pm)^{1+i\lambda}} \cr &=& -\frac{i\lambda}{\pi}\int d^2 w\frac{1}{|w-z|^{2(1+i\lambda)}}\phi^{-\lambda,\pm}_{q(w)}(X).
\eea
Here the shadow transform is normalized to square to the identity. 
In terms of  the symplectic product \be (\phi_1,\phi_2)=\int d^3\Sigma^\mu~\left[\phi_1\partial_\mu\phi_2-\phi_2\partial_\mu\phi_1\right],\ee  modes of the quantum field operator 
$\Phi(X)$
 are  defined by $e.g.$
 \be \Phi^{\lambda,+ }_q =(\Phi,  \phi^{\lambda,+ }_q).\ee
 The quantum commutators for the modes follow from the symplectic products as 
\begin{align}
[\Phi^{\lambda,-}_q,\Phi^{\lambda',+}_{q'}]
= 8\pi^4\delta(\lambda+\lambda') \delta^2(z-z').
\end{align}
 We have the $PT$ relation ($PT(X)=-X$)  \begin{align}\label{dpf}
\phi^{\lambda,+}_q(-X) &= \phi^{\lambda,-}_q(X) \\
\widetilde{\phi}^{\lambda,+}_q(-X) &= \widetilde{\phi}^{\lambda,-}_q(X)
\end{align}

\section{$\a$-vacua}
In this section we construct a one-complex-parameter family of $\a$-vacua  in $M_4$  which reduce to the familiar $\a$-vacua for observers constrained to a hyperbolic $dS_3$ slice. 

The scalar field operator $\Phi(X)$ can be split into a positive frequency and a negative frequency part
\be \Phi(X) = \Phi^+(X)+\Phi^-(X), \ee
where $\Phi^-(X)$ ($\Phi^+(X)$) can be expanded in modes which are negative (positive) frequency with respect to $X$. The standard Poincare-invariant Minkowski vacuum is defined by the condition
\be \label{Pvac}\Phi^-(X)|0_P\rangle=0. \ee
Let $\alpha$ be any complex number with $\text{Re }\alpha <0$. We define the $\a$-vacua by the condition\footnote{Our convention slightly differs from \cite{Bousso:2001mw}, where we have $e^\a$ as the relative factor instead of their $-e^{\a^*}$.}
\be\label{alp} \big(\Phi^-(X)+e^{\a} \Phi^-(-X)\big)|\a \rangle=0. \ee 
$\Phi^-(-X)$ here has an expansion in negative frequency modes of the time reversed $-X$,  hence may be reexpressed in terms of positive frequency components of $\Phi(X)$ as in \eqref{dpf}.

Following roughly the notation in  \cite{Bousso:2001mw} we  define 
\be \label{c and omega} c(\a)\equiv -\frac{1}{2}\left(\ln \tanh \frac{-\text{Re }\a}{2}\right)e^{i \text{Im }\a}, ~~~~ \Omega^{\pm}\equiv {i\over 2}\left(\Phi^{\pm}, PT\Phi\right), ~~~~ U_\a\equiv e^{ c\Omega^++c^*\Omega^-},\ee
where $PT\Phi(X)\equiv\Phi(-X)$ and $U_\a$ is unitary. 
One can show:
\be\left[\Phi^+(X), \Omega^-\right]= \Phi^+(- X),~~~~ \left[\Phi^-(X), \Omega^+\right]= \Phi^-(- X) \ee
from which it follows that 
\be \label{uas} U_\a^{-1} \Phi^-(X) U_\a =N_\a\left(\Phi^-(X)-e^{\a} \Phi^-(-X)\right),\ee
where 
\be
N_\alpha = \frac{1}{\sqrt{1-e^{\alpha+\alpha^*}}}.
\ee
Noting that $\Phi^{\mp}(\pm X)|0_P\rangle=0$ one finds that 
the  squeezed excitation of the Poincare vacuum $|0_P\rangle$
\be \label{ssx} |\a \rangle=U_\a|0_P\rangle\ee
obeys  \eqref{alp}. Upon normal ordering $U_{\a}$ one can show that \eqref{ssx} may alternatively be expressed as $|\a \rangle=C \exp( e^{\a} \Omega^+)|0_P \rangle $ with $C$ a formally vanishing constant.
In terms of the conformal primary wave functions, this vacuum can also be defined by 
\be\label{salp} \big(\Phi_q^{\l-}-e^\a  \Phi_q^{\l+}\big)|\a \rangle=0. \ee

Using \eqref{uas} and \eqref{ssx} the two point function may be expressed
\be \label{grnfn} \langle \a |\Phi(X)\Phi(Y) |\a \rangle= N_{\alpha}^2[G(X,Y)-e^{\a}G(X,-Y)-e^{\a^*}G(-X,Y)+e^{\a+\a^*}G(-X,-Y)]  \ee
where \be G(X,Y)\equiv \langle 0_P |\Phi(X)\Phi(Y) |0_P\rangle={1 \over 4 \pi^2(X_+-Y_-)^2}.\ee 
is the usual Poincare vacuum Wightman function. This relation implies that Green functions  in the $\a$-vacua are related to $\a$-dependent linear combinations of Green functions  in the Poincare vacuum with antipodally reflected external  particle trajectories. Because the antipodal map $X \to -X$ is invariant under Lorentz transformations but not translations, correlation functions in these vacua will possess only Lorentz symmetry and not translation invariance. 

The Green function above naively becomes singular when $e^{\alpha} \to \pm 1$. In this case, the vacua are annihilated by a hermitian operator and become non-normalizable plane wave type states. They can still be defined with some  care  \cite{Ng:2012xp} and have properties which make them of special interest. 

We now show that under dimensional reduction from $M_4$ to $dS_3$, these vacua reduce to the well known de Sitter $\a$-vacua. The flat metric in region $S_4$ of $M_4$ which is spacelike-separated from the origin can be written as a foliation. For $X=(\rho,x)$
\be ds_4^2 = d\rho^2+\rho^2 g_{ab} dx^adx^b,\ee
where $\rho^2=X^2$, $a,b=0,1,2$, and $g_{ab}$ is the metric on the unit radius $dS_3$. Solutions of the 4D massless wave equation can be expanded in the basis
\be \Phi(X)=\int_{-\infty}^{\infty} d \l \Phi^{\lambda}(x)\rho^{-1-i\lambda}.\ee
 The $\Phi^\l$ then represent a continuous infinity of scalar fields on $dS_3$, with 3D masses $m^2=1+\l^2$ \cite{deBoer:2003vf}. The 4D condition \eqref{alp} then implies that the 3D vacuum $|\a_3\rangle$ reduced from $|\a\rangle$ associated with each of these massive fields obeys 
\be\label{alsp} \big(\Phi^{\l-}(x)+e^{\a} \Phi^{\l -}(x_A)\big)|\a_3 \rangle=0, \ee
for every $\l$.\footnote{We discuss below vacua in which $\a$ becomes a function of $\l$.} Here $x_A$ is the anitpodal point to $x$ in $dS_3$, as induced by the $PT$ map $X\to -X$, when embedded in $M_4$. Using $\Phi^{\l -}(x_A)=\Phi^{\l +}(x)$, \eqref{alsp} defines  the familiar $\a$-vacua in dS$_3$. Note in particular that if we take $e^\a\to 0$ we find that the Poincare invariant Minkowski vacuum reduces to the Euclidean $dS_3$ vacuum. This is equivalent to the statement that an observer on a $dS_3$ geodesic uplifts a uniformly accelerated $M_4$ observer who will detect Unruh radiation. 

Although this paper focuses on  a bulk analysis, we note that the bulk operator $(\Phi,PT \Phi)$,  as noted in the de Sitter context in section 4.2 of
\cite{Bousso:2001mw}, is related to a marginal boundary operator constructed from a pair of conformal primary operators, schematically of the form $\int d\lambda \Phi^\lambda \widetilde{\Phi}^{-\lambda}$ or $\int d\lambda \Phi^\lambda\Phi^{-\lambda}$,  which induces a marginal deformation of the boundary CCFT \cite{kapec:2023}. Hence the $\a$-vacua scattering amplitudes may be related to correlators in the deformed CCFT.

\section{$\ta$-vacua}

In this section we construct a one-complex-parameter family of $\ta$-vacua  in Minkowski space which mix up positive and negative  frequency conformal modes with their negative and positive frequency shadow partners. 

The construction of the $\a$-vacua relied on the existence of the Lorentz invariant operator $PT$ obeying $(PT)^2$=1. In this section we similarly construct a second family of Lorentz invariant $\ta$-vacua using the shadow operator $S$ combined with $PT$. This commutes with Lorentz transformations and obeys $(SPT)^2=1$. 

An elegant description of $S$, exploiting $O(4,2)$ conformal invariance of the scalar wave equation, was recently obtained in \cite{Chen:2023tvj,Jorstad:2023ajr} (we review the embedding geometry and relevant transformations in Appendix \ref{appendA}). $O(4,2)$ acts on the $S^3\times S^1$ temporal quotient of the Einstein cylinder, which we denote $\E $. Let $I\in O(4,2) $  denote a  reflection of the $S^3$, obeying $I^2=1$. $I$ breaks the conformal group to  a Poincare plus dilational subgroup. This subgroup preserves a pair of Minkowski diamonds  which sew together to form  $\E$. $I$ exchanges the origin of each diamond with spacelike infinity, and maps an  entire $M_4$ diamond to an inverted and rotated region $IM_4$. 
In \cite{Chen:2023tvj,Jorstad:2023ajr}  it was shown that the analytic continuation of any  solution of the massless scalar wave equation on a diamond $M_4$ to $\E $ obeys:
\be \phi(X)|_{IM_4}=S\phi(X)|_{M_4}, \ee 
$i.e.$ the continued solution on  $IM_4$ equals the shadow of the solution in the original diamond $M_4$. 
Alternately introducing  global coordinates $Z$ on $\E$ we may write 
\be S\phi(Z)=\phi(Z_I), \ee
where $Z_I$ is the image of $Z$ under $I$. 

Repeating all the steps of the previous section, we define the $\ta$-vacua (we again limit ourselves to $ \text{Re }\ta <0$)
\be |\ta \rangle=\tilde U_\ta|0_P\rangle, \ee
where
\be \tilde \Omega^\pm= {i\over 2}(\Phi^\pm,SPT\Phi),~~~~\tilde U_\ta= e^{c(\ta) \Omega^++ c^*(\ta) \Omega^-}. \ee

One finds 
\be\label{salp} \big(\Phi^-(X)+e^\ta \tilde \Phi^-(-X)\big)|\ta \rangle=0 \ee
and 
\be \tilde U^{-1}_\ta \Phi^-(X) \tilde U_\ta =N_\ta(\Phi^-(X)-e^{\ta} \tilde \Phi^-(-X)).\ee
In a conformal basis, this vacuum can be defined by requiring
\be\label{salp} \big(\Phi_q^{\l-}-e^\ta \tilde \Phi_q^{\l+}\big)|\ta \rangle=0. \ee


Using the relation between shadows and bulk conformal inversions, we  sketch the computation of  $\ta$-vacua Green functions. Let  $\Omega^2(Z)$ be the  conformal transformation which maps the $SO(4)\times U(1)$-invariant metric on $\E$ to the flat metric on $M_4$ to   
\be ds_{M_4}^2(Z)=\Omega^2(Z)ds^2_{S^3\times S^1},\ee where $Z=Z(X)$ in the $M_4$ diamond \cite{Jorstad:2023ajr}.

Define \be G_\E(Z,Z')=\langle 0_P|\Phi(Z)\Phi(Z')| 0_P\rangle,\ee
where $| 0_P\rangle$ is the $SO(4,2)$-invariant  extension of the Poincare vacuum form $M_4$ to  
$\E$. Then the conformal transformation properties of $\Phi$ imply that in the flat $X$ coordinates
\be G_\ta(X ,X' ) =\frac{N_{\ta}^2}{\Omega(Z)\Omega(Z')}[G_\E (Z,Z')-e^\ta G_\E (Z,Z_{PTI}')-e^{\ta^*}G_\E (Z_{PTI},Z')+e^{\ta+\ta^*}G_\E (Z_{PTI},Z_{PTI}')].\ee 
where on the right hand side $Z=Z(X)$ and $Z'=Z'(X')$ and  $PTI$ reflects components of $Z$ as detailed in the appendix. This relation implies that Green functions in the $\ta$-vacua are related to $\ta$-dependent linear combinations of Green functions in the Poincare vacuum involving shadow transformed  wavefunctions.

Let's now consider the $dS_3$ reduction of the $\ta$-vacua.  Instead of \eqref{alsp} we get 
\be\label{als} \big(\Phi^{\l-}(x)+e^\ta \tilde \Phi^{\l +}(x)\big)|\ta_3 \rangle=0. \ee
The $M_4\to dS_3$ reduction gives a pair of fields $\Phi^{\pm \lambda}$ with degenerate masses $m^2=1+\lambda^2$. $S$ constructs $\tilde \Phi^\l $ from a non local integral of $\Phi^{-\l}$. Hence the vacuum \eqref{als} is a generalization of the $dS$ $\a$-vacua which mixes degenerate fields. 

\section{Dilation weight dependent vacua}
The defining condition \eqref{alp} for the $\a$-vacua may be modified to 
    \be\label{lp} \big(\Phi^-(X)+e^{\a_0-\a_1 {D}} \Phi^-(-X)e^{\a_1 {D}}\big)|\a_0,\a_1 \rangle=0. \ee 
Here the dilation operator $D$, a generator of $SO(4,2)$,  obeys for constant $a$ 
\be [D,\Phi(X)]=i(\Phi(X)+X^\mu\p_\mu\Phi(X)),~~~~~ e^{ia D}\Phi(X)e^{-ia D}=e^{-a}\Phi(e^{-a}X).\ee 
This modified condition preserves Lorentz invariance because $D$ commutes with $SO(3,1)$ Lorentz generators.  Note that $e^{iaD}$ is in $SO(4,2)$ only when $a$ is real. 

Using the formula for conformal primary modes of the field operator,  dilations act by
\be  e^{-iaD}\Phi_q^{\l\pm}e^{ia D}=e^{ia\l}\Phi^{\l\pm}_q.\ee
In this basis we may 
write\footnote{In fact the distinction between positive and negative frequency in \eqref{lp} has some ambiguity for real $\a$ which is complex dilations. This ambiguity can be fixed by viewing \eqref{lzp} as the definition of \eqref{lp}.}
\be\label{lzp} \big(\Phi^{\l-}_q-e^{\a_0+\a_1\l} \Phi^{\l +}_q\big)|\a_0,\a_1\rangle=0. \ee
In this case the vacuum state associated with each $dS_3$ field $\Phi^{\l}(x), ~\l \in {\mathbb R} $,  has a different  $\a$ parameter.

\section{The hyperbolic Rindler vacuum}

In \cite{Bousso:2001mw} it was noticed that for $dS_3 $ there is a special choice of $(\a_0,\a_1)=(i\pi,-\pi)$
for which the in/out Bogolyubov transformation becomes trivial and there is no particle production. The uplift of this vacuum to $M_4$, which we refer to as the hyperbolic Rindler vacuum $|H\rangle$, is defined by \footnote{We note that $e^{-\pi D}X^\mu e^{\pi D}=-X^\mu$  and  $e^{-\pi D}\Phi^{\l -}_qe^{\pi D}=e^{-\pi \l}\Phi^{\l +}_q$. This differs from what might have been expected from \eqref{dpf} because the rotation of $X^\mu$ through the complex plane crosses a branch cut, and enables us to write \eqref{alf} as $\big(\Phi^{\l-}_q+e^{-\pi D}\Phi^{\l -}_qe^{\pi D}\big)|H\rangle=0$. This in turn suggests a relation with  the special $\a$-vacua with $e^\a=\pm 1$.}
\be\label{alf} \big(\Phi^{\l-}_q+e^{-\pi \l}\Phi^{\l +}_q\big)|H\rangle=0. \ee
Interestingly,  the wave function associated to these field modes takes the simple form 
\be \psi^\l_q(X)\equiv \phi^{\l-}_q(X)+e^{-\pi \l}\phi^{\l +}_q(X) = -2ie^{-\pi\lambda/2}\sinh\pi\lambda\Gamma(1+i\lambda){\Theta(q \cdot X) \over (q\cdot X)^{1+i\l}}.\ee
These are Rindler-like modes which vanish above the null plane $q\cdot X=0$. Hence  $H$ is a hyperbolic version of the Rindler vacuum. The usual Rindler vacuum is defined by taking annihilation operators to be those with negative frequency  with respect to the proper time of observers who are uniformly accelerated in a particular direction. The hyperbolic Rindler vacuum instead employs the proper time of  observers who are uniformly radially accelerated away from the origin.

The special simple properties and symmetries of the vacuum $|H\rangle$ suggest that it, along with  the $\a$-vacua with $e^\a=\pm 1$,  may  provide  a natural home  for celestial holography. For Vasiliev gravity in $dS_4$, there is a related pair of special vacua whose extrapolated Green functions equal those of the free or interacting $SP(N)$ CFT. These are the natural states in which to define $dS_4/CFT_3$, and the dictionary is more complicated for  other bulk vacua. These special hyperbolic vacua  may play a similar role for celestial holography. 

One may further consider Lorentz invariant vacua in which $\a$ is an arbitrary function of $\l$ in \eqref{lzp}. Another interesting case is the vacuum annihilated by all $\Phi^{\pm \lambda}_q$ with positive $\lambda$\cite{cmsw}. While we expect that a classification of all Lorentz-invariant vacua should be possible, we have not attempted to give one here.  
\section*{Acknowledgements}

We are grateful to Jordan Cotler, Daniel Kapec, Noah Miller, Zixia Wei and especially Atul Sharma for useful discussions. This work was supported by DOE grant de-sc/0007870,   NSF GRFP grant DGE1745303 and the Simons Collaboration on Celestial Holography. 

\appendix
\section{Geometry of Embedding Space and Inversions} \label{appendA}
In this appendix we review the structure of inversions in the embedding space of $\mathbb{R}^{1,3}$. The embedding space for $\mathbb{R}^{1,3}$ is $\mathbb{R}^{2,4}$, with coordinates $Z = (Z^{-1},Z^I, Z^4)$, $I = 0,\ldots,3$. Then, the quotient of the lightcone
\begin{equation}
    (Z^{-1})^2 + (Z^0)^2 - (Z^1)^2-(Z^2)^2-(Z^3)^2-(Z^4)^2 = 0
\end{equation}
by positive rescalings $Z \to tZ$ gives Penrose's conformal compactification of $\mathbb{R}^{1,3}$ \cite{Jorstad:2023ajr}. A standard embedding of $\mathbb{R}^{1,3}$ is
\begin{equation}
    \left(Z^{-1},Z^\mu,Z^4\right) = \pm \left(\frac{1+|X|^2}{2},X^\mu,\frac{1-|X|^2}{2}\right)
\end{equation}
up to positive rescalings. The Minkowski metric is the pullback of the weight-0 structure
\begin{equation}
    ds^2_{M_4} = \Omega(Z)^2\eta_{IJ}dZ^IdZ^J,
\end{equation}
where $\Omega(Z) = (Z^{-1} + Z^4)^{-1}$.

We can pull these coordinates to the temporal quotient of the Einstein cylinder $S^1 \times S^3$ using the embedding \cite{Jorstad:2023ajr}
\begin{equation}
    Z^I = (\cos\tau,\sin\tau,\sin\pi\sin\theta\cos\phi,\sin\psi\sin\theta\sin\phi,\sin\psi\cos\theta,\cos\psi).
\end{equation}
In these coordinates, the Minkowski metric takes the form 
\begin{equation}
    ds^2_{M_4} = \Omega^2ds^2_{S^1\times S^3},
\end{equation}
where $\Omega = (\cos\tau+\cos\psi)^{-1}$.

Given a field $\Phi(X)$ in Minkowski space of weight 1, we can uplift it to a homogeneous function of the lightcone 
\begin{equation}
    \Omega(Z)\Phi(X(Z))= \Phi(Z).
\end{equation}
$\Phi(Z)$ transforms as a scalar under conformal transformations of $\mathbb{R}^{1,3}$ and has weight $-1$ under the transformation $Z \to tZ$. Using this expression, we have that the uplift of the propagator is
\begin{equation}
    G(Z,Z') = \Omega(Z)\Omega(Z')G(X,X').
\end{equation}
This transforms as a scalar under $\mathbb{R}^{1,3}$ conformal transformations \cite{Jorstad:2023ajr}.

In terms of the projective coordinates $Z$, conformal inversions $X^\mu \to \frac{X^\mu}{|X|^2}$ take the simple form 
\begin{equation}
    Z \to Z_I = (Z^{-1},Z^0,Z^1,Z^2,Z^3,-Z^4)
\end{equation}
In terms of the coordinates on $S^1 \times S^3$, this reflection acts by sending $\psi \to \pi-\psi$. The combination $PTI$ acts on $Z$ by 
\begin{equation}
    Z \to Z_{PTI} = (Z^{-1},-Z^0,-Z^1,-Z^2,-Z^3,-Z^4)
\end{equation}

\bibliography{hvac.bib}
\bibliographystyle{utphys}

\end{document}